\documentclass[twocolumn]{article}
\usepackage[sorting=none]{biblatex} 
\begin{title}
    \large
    \title{Review of Cookie Synchronization Detection Methods}
    \small 
    \vspace{0.1cm}
    \author{Jake Smith \\ \emph{University of California, Davis} \\ Email: jsssmit@ucdavis.edu}
\end{title}

\begin{document}

\maketitle

\section{Abstract}
The research community has deemed cookie synchronization detection an inherently challenging task \cite{solomos2019clash,englehardt2016online,agar2014web}.
  Studies aiming to identify cookie synchronizations often share high-level design choices, but deviate amongst low-level implementations.        
  For example, the majority of studies label a cookie synchronization iff a user identifier is shared with a third party; however, there is a lack of consistency among implementations, such as party relations or identifier value definitions, or whether such definitions are even included.
  This review intends to provide a record of established methods and promote standardization of methods choice in future work.
  \newline \newline 
  \textbf{CCS Concepts:} Web protocol security; Network privacy and anonymity; Surveillance. \newline
  \textbf{Keywords:} cookie synchronization; cookie matching; tracking; cookies; methods.

\section{Introduction}
The sharing of user browsing information is necessary for the Internet advertising and tracking industries to serve targeted ads \cite{bashir2016tracing,ghosh2015match,olejnik_tran_castelluccia_2014,pachilakis2021youradvalue}, perform cross-device tracking \cite{brookman2017cross}, and sell user information \cite{olejnik_tran_castelluccia_2014,pachilakis2021youradvalue,papadogiannakis2021user}. 
Browser cookies are a standard container for user browsing data, and the sharing of first party cookies with third parties is restricted by the Same-Origin policy \cite{same_origin_policy} to protect user privacy \cite{papadogiannakis2021user, papadopoulos2018exclusive,279944}. 
Cookie synchronization is used to bypass the Same-Origin policy and share first party cookies with third parties to support the advertising and tracking ecosystem  \cite{papadogiannakis2021user, papadopoulos2018exclusive,papadopoulos2019cookie}.
Cookie synchronization is defined by a variety of terms in the research community, such as cookie matching, cookie linking, cookie leaking, and ID syncing.

\section{Background}
\subsection{Browser Cookies and User Identifiers}
Cookies are \texttt{key=value} pairs set on a user's browser to bring state to the HTTP protocol and provide session management, user personalization, and tracking functionality. 

Browser cookies can be set by the \texttt{Set-Cookie} header of HTTP responses \cite{279944,falahrastegar2016tracking,mdn_http_setcookie} or the \texttt{document.cookie} operation of JavaScript embedded in a visited website \cite{doc.cookie}.

Cookie synchronization involves the sharing of cookie values that can uniquely identify a user (i.e. the cookie value is unique to one user).
This review defines such cookie values as \textsl{identifiers}. 
Methods used to define and label identifiers are discussed in \textsl{Section 7.2}. 

\subsection{Party Relations}
First party cookies are set by a user requested domain, and third party cookies are set by an entity (i.e. domain or parent organization) other than the domain requested.

\subsection{Cookie Synchronization}
Cookie synchronization is defined as the sharing of a first or third party identifier with another third party, which can be initiated by an embedded third party resource, third party redirect, or the first party itself \cite{papadopoulos2019cookie,fouad2018missed,papadogiannakis2021user}.   

\subsection{How is Cookie Synchronization Performed?}
Assume a user is browsing \texttt{website1.com} and \texttt{website2.com}, and there exists tracking entities \texttt{tracker1.com} and \texttt{tracker2.com} who both set identifiers on the user’s browser, \texttt{ABC} and \texttt{123}, respectively. 
The user later visits \texttt{website3.com}, which has an embedded resource from \texttt{tracker1.com} that initiates a GET request to \texttt{tracker1.com}. 
\texttt{tracker1.com} responds with a 3XX redirect instructing the user’s browser to issue another request to \texttt{tracker2.com}, with the identifier for \texttt{tracker1.com} (\texttt{ABC}), placed in the parameters of the requested URL\footnote{Additional locations to share identifiers are discussed in \textsl{Section 7.2}.}. 
\texttt{tracker2.com} is now able to link its identifier (\texttt{123}) with \texttt{tracker1.com}’s identifier (\texttt{ABC}) \cite{papadogiannakis2021user, papadopoulos2018exclusive,papadopoulos2019cookie}. \par
\subsection{User Privacy Erosion}
Cookie synchronization allows a third party to reconstruct portions of a user’s browsing history by receiving the visited first party site in the \texttt{Referer} field of a GET request header \cite{papadopoulos2019cookie,mdn_http_referer}. 
Websites visited over TLS are not exempt from this history leakage, as  plaintext HTTP requests to third parties share URLs requested using HTTPS \cite{papadopoulos2018exclusive,papadopoulos2019cookie,sanchez2021journey}. 
As a tracker learns more third party identifiers for a single user, it can reconstruct a larger portion of her browsing history \cite{papadopoulos2019cookie}. \par
Cookie respawning methods such as \texttt{evercookie} \cite{kamkar_2010} can enable third parties to re-identify users after clearing browser cookies. A respawned identifier can be re-synced with a tracker, effectively eliminating a user’s ability to delete browser cookies.
This enables third parties to track users and join browsing histories across browser refreshes \cite{papadopoulos2019cookie,agar2014web}. \par
Server-to-server user data merges are facilitated by cookie synchronization. Separate tracker data-sets of known user information can be combined by linking respective identifiers for each tracker \cite{papadopoulos2019cookie,agar2014web}.
\subsection{Cookie Synchronization and the Advertising Industry}
Advertising companies are motivated to collect as much user information as possible in order to serve the most targeted ads; Bashir et. al. \cite{bashir2016tracing} report Demand-Side Platforms (DSPs) place higher bids to serve users whom they have more information about. 
Cookie synchronization enables this information acquisition by sharing user browsing data and linking tracker databases, which enables ad targeting based on web history \cite{papadopoulos2019cookie}.   
Papadopoulos et. al. \cite{papadopoulos2019cookie} report ad related domains are the most prevalent entities involved in cookie synchronization, participating in 75\% of all synchronizations and acquiring as much as 90\% of all identifiers synced.  
\section{Related Work}
As early as 2014, Olejnik et. al. \cite{olejnik_tran_castelluccia_2014} showed how advertisers use cookie synchronization in real-time bidding (RTB) to reconstruct and share browsing history.
HTTP traffic and browser cookies were collected from 100 real users browsing more than 70 sites each. 
After 70 site visits, a user experienced on average 100 cookie synchronizations with 30 domains involved. 
Acar et. al. \cite{agar2014web} investigated the effect Firefox’s privacy settings \{\texttt{Allow Third Party Cookies}, \texttt{Allow All Cookies but Do Not Track}, \texttt{Block Third-Party Cookies}\} have on the number of cookie synchronization a user encounters. 
Multiple crawls of the Alexa top 3,000 sites were performed with browser cookies logged. 
When third party cookies were allowed, 596 identifiers were synced over 407 unique first parties, with 323 third parties involved. 
Selecting \texttt{Do Not Track} only decreased the number of domains involved in cookie synchronization by 2.9\% and identifiers shared by 2.6\%. 
When third party cookies were blocked, this decreased the number of identifiers synced to 353 over 321 first parties, with 129 third parties involved. 
They report 3 instances of respawned cookies being synced over two 3,000 site crawls.

Papadopoulos et. al. \cite{papadopoulos2019cookie} investigated the prevalence of cookie synchronization events in mobile web traffic. 
The study collected 850 mobile users HTTP traffic for 12 months. 
263,635 cookie synchronizations were detected over 179M total requests, with 22,329 identifiers shared; 91.996\% of the shared identifiers were located in URL parameters, 3.705\% in the \texttt{Referer} URL, and 3.771\% in the URL path. 
The study reports 5\% of identifiers set in TLS sessions being leaked over plain HTTP, as well as the websites visited in the \texttt{Referer} field. 

Brookman et. al. \cite{brookman2017cross} examined the extent of cross-device tracking visible to an end-user, including cookie synchronizations. 
The study crawled the Alexa top 100 websites four times each.
They report 106 unique third parties syncing identifiers with 210 other third parties. 

Englehardt et. al. \cite{englehardt2016online} performed an extensive analysis of online tracking using their open source crawler, OpenWPM. 
They collected web traffic and browser cookies from two crawls of the top 10K Alexa websites. 
They report the majority of common third parties embedded in websites participating in cookie synchronization: 45 of the top 50, 85 of the top 100, and 157 of the top 200.

Papadopoulos et. al. \cite{papadopoulos2018exclusive} investigated TLS privacy breaches facilitated by cookie synchronization, specifically the sharing of websites visited and identifiers set over HTTPS. 
The top 12K Alexa websites were crawled, with 440K HTTP(S) requests logged.
They report 89,479 HTTP(S) syncing requests (i.e. HTTP redirects sharing an identifier) occurring from 32\% of the crawled domains; 17,171 unique identifiers were shared with 733 unique domains.
Of the 8,398 websites visited over TLS, 2,317 websites were involved in cookie synchronization. 
Most critically, these TLS websites conducted 2,879 cookie synchronizations with non-TLS websites and leaked 174 HTTPS visits over plaintext. 
They report 1 in 13 TLS-supported websites performing cookie synchronization over HTTP. 

Urban et. al. \cite{urban2020measuring} performed a longitudinal study documenting the effects of the General Data Protection Regulation (GDPR) on cookie synchronization rates in the European Union (EU). 
12 measurements were performed, with one occurring a month before the GDPR going into effect (May 2018), and the rest performed each month after.
Each measurement instrumented 400 individual browsing profiles (i.e. unique browsing instances).
The measurements each crawled an average of 8.5K domains, totalling over 2.5M requests over the year. 
After the legislation's passing in May 2018, they report an immediate drop in the number of cookie synchronizations per month ($\sim$510) in relation to the pre-GDPR measurement (898); a year later, this number decreased to $\sim$480 cookie synchronization per month. 
The number of third parties conducting cookie synchronizations per month also decreased from $\sim$12K to $\sim$10.2K. The number of involved third parties per month gradually recovered over the year to $\sim$12K. 
The study claims “cookie synchronization is still used in practice, but its extent is significantly reduced and still declining” in the EU \cite{urban2020measuring}. 
This claim is not supported by the results of later studies conducted in the EU by Fouad et. al. \cite{fouad2018missed} and Papadogiannakis et. al \cite{papadogiannakis2021user}.

Fouad et. al. \cite{fouad2018missed} investigated the role of 1x1 pixel images and other embedded content types in initiating cookie synchronization.
They conducted two crawls of the Alexa top 10k domains, and successfully crawled 8,744 domains.
They report 34.36\% of tracking was initiated by scripts, 23.34\% by pixels, 20.01\% by text/html, 8.54\% by large images, and 4.32\% by application or JSON. 
Of the 8,744 websites crawled, 67.96\% were involved in cookie synchronization, with 17,425 third parties involved.
Third party identifiers were shared with other third parties in 22.73\% of websites with 1,263 unique partners.

Sanchez-Rola et. al. \cite{sanchez2021journey} conducted a large scale crawl of the Tranco top 1M most accessed domains list to reconstruct the cookie ecosystem, clarifying known roles and defining novel ones involved in the creation and sharing of cookies. 
They define the ghost cookie, which is created by an embedded third party script on a first party website that sets a first party cookie. 
The study claims the existence of a ghosted cookie decreases a first party's control over the cookies their web-page sets on a browser.
They report 8.97M cookie synchronization across 387K websites, with the most common sender and receiver relationship (48\%) being the own sender to own receiver (i.e. a first party ghost cookies shared with the third party who embedded the script). 
52.4\% of domains experience at least one cookie synchronization or cookie value overwriting event.
Reflecting the results of Papadopoulos et. al. \cite{papadopoulos2018exclusive,papadopoulos2019cookie}, 37.71\% of cookies synchronized over HTTP were created in a TLS session. 

Papadogiannakis et. al. \cite{papadogiannakis2021user} investigated whether third party trackers respect cookie consent banner choices \{\texttt{No Action}, \texttt{Reject All Cookies}, \texttt{Accept All Cookies}\}. 
Their data-set was derived from the Tranco top 850K sites and successfully crawled 27,953 domains containing a Consent Management Platform (CMP). 
They specify two types of cookie synchronization relationships. 
They define a \texttt{First-Party ID Leak} if a first party identifier is shared with a third party, and a \texttt{Third-Party ID Synchronization} if a third party identifier is shared with a third party. 
When the user takes \texttt{No Action}, 52.88\% and 24.03\% of websites conduct \texttt{First-Party ID Leaking} and \texttt{Third-Party ID Synchronization}, respectively. When \texttt{Rejecting All Cookies}, 56.41\% and 26.20\% of websites conduct \texttt{First-Party ID Leaking} and \texttt{Third-Party ID Synchronization}, respectively.

\section{Purpose}
This review intends to document the variety of methods employed to detect cookie synchronization. 
All studies under review must log HTTP data and label cookie synchronizations from the collected network traffic.

\section{Data-set Collection Methods}
\textbf{Crawled Data-set:}
Web crawlers instrumented include OpenWPM \cite{brookman2017cross,279944,urban2020measuring,fouad2018missed,englehardt2016online,solomos2019clash}, Chromium-based crawlers \cite{bashir2016tracing,sanchez2021journey,bashir2018diffusion,bidelman_2018}, Selenium-based crawlers \cite{papadopoulos2018exclusive,agar2014web,selenium}, or custom crawlers \cite{papadogiannakis2021user}.

\textbf{User Data Collection:} To collect the HTTP traffic of real users, study-specific browser plugins are installed on a user's browser \cite{olejnik_tran_castelluccia_2014,pachilakis2021youradvalue,papadopoulos2019cookie,falahrastegar2016tracking}.

Henceforth, the term \emph{user} will refer to the browser instance instrumented, regardless of whether the study collected crawled or real user data.
\section{Labeling Cookie Synchronizations by Shared Identifiers}
\subsection{Shared Identifier Heuristic}
The majority of cookie synchronization detection methods draw inspiration from the shared identifier heuristic proposed by Olejnik et. al. \cite{olejnik_tran_castelluccia_2014}. 
This method labels a cookie synchronization iff an identifier is shared in a HTTP request's URL parameters to an entity other than the entity who set the cookie (i.e. a third party) \cite{olejnik_tran_castelluccia_2014,pachilakis2021youradvalue,fouad2018missed,falahrastegar2016tracking}. 
An entity can be defined as either a domain or the parent organization of a domain. \par
Related methods build on this heuristic by additionally extracting identifiers shared with third parties from the URL path of requests \cite{papadogiannakis2021user,papadopoulos2019cookie,agarwal2020stop}, \texttt{Referer} URL of requests\footnote{As of November 2020, the HTTP \texttt{Referrer-Policy} default directive has been updated to \texttt{strict-origin-when-cross-origin} to only share the origin of a request. This prevents identifiers from being shared in the path and querystring \cite{referrerpolicy}.}  \cite{papadogiannakis2021user, papadopoulos2018exclusive,papadopoulos2019cookie,englehardt2016online,agarwal2020stop,agar2014web}, redirect \texttt{Location} URL \cite{agar2014web}, nonstandard request and redirect headers \cite{279944}, or POST request bodies \cite{papadogiannakis2021user}. 
\subsection{Extracting Identifiers from Browser Cookies}
\subsubsection{What Defines an Identifier?}
A cookie set on a user’s browser is an identifier iff the cookie’s value can identify a specific user (i.e. the value is mapped to only one user). 
These identifying  cookie values and the entities who set them are stored to later detect instances of identifiers shared in HTTP traffic.
This method confirms that a cookie value shared with a third party can uniquely identify the user who initiated the third party request \cite{pachilakis2021youradvalue,brookman2017cross,papadogiannakis2021user,papadopoulos2018exclusive,279944,papadopoulos2019cookie,agar2014web,urban2020measuring,fouad2018missed,sanchez2021journey,englehardt2016online,falahrastegar2016tracking}. 
\subsubsection{Extracting Browser Cookies}
To create the set of all cookies set on a user’s browser, cookie values are extracted from the \texttt{Set-Cookie} header of HTTP responses \cite{279944,falahrastegar2016tracking,mdn_http_setcookie,papadopoulos2019cookie} or \texttt{Cookie} header of HTTP requests \cite{279944,mdn_http_cookie}. \par 
Solomos et. al. \cite{solomos2019clash} use OpenWPM’s \texttt{javascript\_instrument} \cite{englehardt2016online} to log cookie values set by JavaScript embedded in visited web pages. 
\subsubsection{User Identifier Filtering}
The following restrictions are used to filter identifier cookie values from the original browser cookie set.
\newline 
\newline
\textbf{Value Length Restrictions:} Identifiers often have minimum length requirements: cookie values $>$ 10 characters \cite{olejnik_tran_castelluccia_2014,pachilakis2021youradvalue,papadopoulos2019cookie,agarwal2020stop}, $>$ 8 characters \cite{279944}, $>$ 7 characters \cite{urban2020measuring,englehardt2016online,falahrastegar2016tracking}, and $>$ 5 characters \cite{papadogiannakis2021user}. 
Of studies that provide identifier length restrictions, only one provides an upper bound: $\leq$ 100 characters \cite{englehardt2016online}. \newline \newline
\textbf{Value Character Quality Restrictions:} 
Identifiers can be extracted based on character values.
Studies that set character value restrictions only extract cookie values consisting of alphanumeric characters and other common characters \cite{englehardt2016online,fouad2018missed,279944}. Common character values include [\texttt{-, \_, =}], with \texttt{=} indicating a \texttt{key=value} pair \cite{englehardt2016online}. 
Fouad et. al. \cite{fouad2018missed} also consider the comma and period and exclude the equals sign.  \newline \newline
\textbf{Delimiter Parsing:} To extract consecutive identifier strings bounded by known characters, cookie values can be parsed (i.e. split) at these common delimiters. 
All studies that split consecutively shared identifiers consider [\texttt{\&, ;}] to be delimiters, except Ghosh et. al. \cite{ghosh2015match} who consider the colon rather than semicolon \cite{papadogiannakis2021user,279944,agar2014web,urban2020measuring,fouad2018missed,englehardt2016online,falahrastegar2016tracking,agarwal2020stop,papadopoulos2019cookie}. \newline \newline
\textbf{Similarity Measurement:} Identifiers can be extracted by uniqueness.
All studies extracting identifiers based on string entropy use the Ratcliff/Obershelp Algorithm \cite{black_2021} with a provided maximum similarity score: eliminate cookie values $>$ 66\% similar to another cookie value \cite{englehardt2016online}, $>$ 33\% similar \cite{brookman2017cross,agar2014web,agarwal2020stop}, or not provided \cite{urban2020measuring}. \newline \newline
\textbf{Multiple Values Set for a \texttt{Key=Value} Pair:} Falahraster et. al. \cite{falahrastegar2016tracking} and Urban et. al. \cite{urban2020measuring} exclude any cookie value extracted from a \texttt{key=value} pair containing more than one value.\newline \newline
\textbf{\texttt{Key=Value} Pairs with Dynamic Values:} Cookie values can be eliminated if the key’s value changes over the course of a crawl or user browsing session \cite{agar2014web,englehardt2016online,agarwal2020stop}. \newline \newline
\textbf{Keyword Filtering:} Papadogiannakis et. al. \cite{papadogiannakis2021user} use a manually curated list of keywords to eliminate cookie values containing dates, timestamps, regions, locale, URLs, prevalent keywords, consent information (e.g. values of the keys \texttt{euconsent}, \texttt{eupubconsent}, \texttt{\_\_cmpconsnent}, \texttt{\_\_cmpiab}), or end in common file extensions. \newline \newline
\textbf{Filtering Non-Unique Strings:} Studies with access to multiple cookie data-sets from multiple crawls or user browsing sessions can eliminate cookie values present for multiple crawls or users \cite{papadopoulos2019cookie,urban2020measuring,fouad2018missed,falahrastegar2016tracking}. \newline \newline
\textbf{Session Cookie Values}: Session cookies are deleted at the end of a browsing session and their values can be eliminated \cite{mdn_http_cookie}. 
Studies that eliminate session cookies examine the \texttt{Expires} and \texttt{Max-Age} attributes \cite{mdn_http_cookie} and eliminate values associated with cookies lacking an expiration date  \cite{papadopoulos2018exclusive,papadopoulos2019cookie} or expire earlier than a specified future date: earlier than 90 days \cite{englehardt2016online} or 30 days \cite{agar2014web}. \newline
\subsection{Detecting Identifiers Shared in HTTP Traffic}
\subsubsection{Labeling Requests to First or Third Parties} 
Studies that label the party relation of (referrer, request) pairs only label identifiers shared in requests to third parties \cite{olejnik_tran_castelluccia_2014,pachilakis2021youradvalue,brookman2017cross,papadogiannakis2021user,papadopoulos2018exclusive,279944,papadopoulos2019cookie,urban2020measuring,fouad2018missed,sanchez2021journey,englehardt2016online,solomos2019clash,falahrastegar2016tracking,agarwal2020stop}.\par
\textbf{Parent Organization Mapping:} Domain names can be mapped to parent organizations using DNS \texttt{whois} records and blacklists \cite{papadopoulos2018exclusive,papadopoulos2019cookie,falahrastegar2016tracking,agarwal2020stop} or the \texttt{WhoTracks.me} database \cite{sanchez2021journey,whotracksme}.
To resolve domain names obfuscated by CNAME cloaking \cite{cointepas2019cname}, Sanchez et. al. \cite{sanchez2021journey} use the NextDNS blocklist \cite{nextDNS} to resolve these cloaked domains to known trackers;
\texttt{tldExtract} \cite{tldextract} is then used to determine the private suffix of each domain;
private suffixes are mapped to parent organizations using the \texttt{Disconnect} \cite{disconnectlist}, \texttt{WhoTracks.me} \cite{whotracksme}, and \texttt{webxray} \cite{webxray} lists. \par 
\textbf{String Matching:} Domain name string matching is also common, with matches indicating a first party and mismatches indicating a third party \cite{olejnik_tran_castelluccia_2014,pachilakis2021youradvalue,brookman2017cross,papadogiannakis2021user}. \par
\textbf{Englehardt et. al. Case Study:} Englehardt et. al. \cite{englehardt2016online} label request party relations using the Mozilla Public Suffix list \cite{publicsuffixlist}; iff the landing page's domain name and public suffix (not including subdomains) do not match a request's domain name and public suffix, the request is labeled as to a third party.

\subsubsection{HTTP Identifier Sharing Locations}
The research community has examined the following HTTP elements for instances of shared identifiers using exact string matching, with matches indicating a cookie synchronization. \newline

\textbf{HTTP GET Requests:} URL query parameters \cite{olejnik_tran_castelluccia_2014,agar2014web,englehardt2016online,brookman2017cross,papadopoulos2018exclusive,fouad2018missed,papadopoulos2019cookie,agarwal2020stop,papadogiannakis2021user,279944}, URL path\footnote{Studies who report examining URLs–without specifying which elements–are assumed to check both the path and querystring.}  
\cite{agar2014web,englehardt2016online,brookman2017cross,papadopoulos2018exclusive,papadopoulos2019cookie,agarwal2020stop,papadogiannakis2021user,279944}, \texttt{Referer} URL\footnote{As of November 2020, the HTTP \texttt{Referrer-Policy} default directive has been updated to \texttt{strict-origin-when-cross-origin} to only share the origin of a request. This prevents identifiers from being shared in the path and querystring \cite{referrerpolicy}.} \cite{englehardt2016online,papadopoulos2018exclusive,papadopoulos2019cookie,agarwal2020stop,papadogiannakis2021user,agar2014web}, and non-standard headers \cite{279944}. \newline

\textbf{HTTP Redirects:} \texttt{Location} URL \cite{agar2014web,englehardt2016online,agarwal2020stop} and non-standard headers \cite{279944}. \newline

\textbf{HTTP POST Requests:} Request bodies \cite{papadogiannakis2021user}.
\subsubsection{Papadopoulos et. al. Shared Identifier Labeling Case Study}
Papadopoulos et. al. \cite{papadopoulos2019cookie, papadopoulos2018exclusive} implemented a distinct method of detecting instances of shared identifiers over two cookie synchronization studies.

Rather than using string matching to label instances of shared identifiers, they first extract all ID-looking strings from GET request URL paths, query parameters, and \texttt{Referer} headers. 
An ID-looking string is defined by the same qualities used for filtering identifiers from browser cookies \{\textsl{Section 7.2}\}.

The study stores detected ID-looking strings in a hashtable with the receiving domain.
If an ID-looking string is seen for the first time in an HTTP element, the string is added to the hashtable with the requested domain.
If an ID-looking string is seen for at least the second time, all requests carrying it are labeled as an ID-sharing event.

Cookie synchronizations are labeled from the ID-sharing event set; iff an ID-looking string present in an ID-sharing event matches a known identifier, the ID-sharing event is labeled a cookie synchronization.

\section{Alternative Cookie Synchronization Detection Methods}
\subsection{Decision Tree Classifier of Encrypted Identifier Synchronization}
Papadopoulos et. al. \cite{papadopoulos2019cookie} trained a decision tree model to detect cookie synchronizations of encrypted identifiers.
The model does not consider the presence of a shared, known identifier when classifying cookie synchronizations.

The study assumes an equal distribution of HTTP traffic feature variability between cookie synchronization of non-encrypted and encrypted identifiers. 
The training and testing sets were labeled by non-encrypted cookie synchronizations detected using the study’s shared identifier heuristic.
The features selected include requested entity name, type of entity \{\texttt{Content, Social, Advertising, Analytics, Other}\}, URL parameter names, location of hashed identifier \{URL parameter, URL path, \texttt{Referer} URL parameter\}, HTTP status code, browser type, and number of parameters.
\subsection{Labeling Cookie Synchronizations in Retargeted Ad Serving Information Flows}
Bashir et. al. \cite{bashir2016tracing} collect the resource inclusion chain for all websites crawled. 
At a high level, a cookie synchronization is labeled iff an auction is held by the \texttt{publisher-side} and requests between the Supply-Side Platforms (SSP) of the chain directly include a resource.\par
The study defines the following terminology.
\texttt{Personas} are individually created to represent 90 unique categories of shoppers by browsing specific products on e-commerce sites. 
These categories are used to later compare with the qualities of retargeted ads for each \texttt{persona}.
A \texttt{publisher-side} resource chain serves a retargeted ad to a user’s browser. 
\texttt{pub} is the root node’s publisher domain. 
\texttt{d} is the last entity in a chain and serves the ad.
\texttt{s} denotes a SSP. 
\texttt{shop} is the e-commerce site domain of the retargeted ad.

Cookie synchronizations are labeled iff \texttt{s} and \texttt{d} are adjacent at the end of a chain, \texttt{d} observes the \texttt{persona} at \texttt{shop}, and a request from \texttt{s} to \texttt{d} (or \texttt{d} to \texttt{s}) is present in a chain prior to the retargeted ad being served \cite{bashir2016tracing}.
\subsection{Labeling Tracker to Tracker Cookie Synchronizations with Pre-Existing Data-sets}Bashir et. al. \cite{bashir2018diffusion} and Solomos et. al. \cite{solomos2019clash} label any (tracker, tracker) referrer-request pair as a cookie synchronization iff the pair is present on a list of known cookie synchronizing third parties \cite{bashir2016tracing,papadopoulos2019cookie}.

\section{Acknowledgements}
The author would like to thank Dr. Zubair Shafiq and Dr. Katie Rodger for their technical and expository insights. 

\printbibliography

\end{document}